\documentclass[twocolumn]{aastex631}

\usepackage{amsmath}
\usepackage{ulem}
\usepackage{xcolor}
\usepackage{amsmath}
\usepackage{cases}
\usepackage[T1]{fontenc}
\usepackage{subfigure}
\usepackage{epsfig}
\usepackage{threeparttable}

\begin{document}
\title{Short-duration GRB 250221A Afterglow Driven by Two-Component Jets from the merger of a compact star}

\author[0009-0001-8132-911X]{Xiao Tian}
\affiliation{Guangxi Key Laboratory for Relativistic Astrophysics, Department of Physics, Guangxi University, Nanning 530004, People’s Republic of China; lhj@gxu.edu.cn}

\author[0000-0001-6396-9386]{Hou-Jun L\"{u}}
\altaffiliation{Corresponding author (LHJ) email: lhj@gxu.edu.cn}
\affiliation{Guangxi Key Laboratory for Relativistic Astrophysics, Department of Physics, Guangxi University, Nanning 530004, People’s Republic of China; lhj@gxu.edu.cn}

\author[0009-0003-6940-2171]{Xiao-Xuan Liu}
\affiliation{Guangxi Key Laboratory for Relativistic Astrophysics, Department of Physics, Guangxi University, Nanning 530004, People’s Republic of China; lhj@gxu.edu.cn}

\author[0009-0000-0467-0050]{Xiao-Fei Dong}
\affiliation{School of Astronomy and Space Science, Nanjing University, Nanjing 210023, People’s Republic of China}

\author[0000-0002-9037-8642]{Jia Ren} 
\affiliation{Purple Mountain Observatory, Chinese Academy of Sciences, Nanjing 210023, People’s Republic of China}

\author[0009-0008-6247-0645]{Wen-Long Zhang} 
\affiliation{Purple Mountain Observatory, Chinese Academy of Sciences, Nanjing 210023, People’s Republic of China}
\affiliation{School of Astronomy and Space Sciences, University of Science and Technology of China, Hefei 230026, People’s Republic of China}

\author[0000-0002-7044-733X]{En-Wei Liang}
\affiliation{Guangxi Key Laboratory for Relativistic Astrophysics, Department of Physics, Guangxi University, Nanning 530004, People’s Republic of China; lhj@gxu.edu.cn}

\begin{abstract}
GRB 250221A is a short gamma-ray burst (GRB) at redshift $z=0.768$, with a duration of 1.8 s and no extended emission in either Swift/Burst Alert Telescope or Konus-Wind bands. A remarkable rebrightening feature in both optical and X-ray bands was observed at $\sim$0.6 days after the burst trigger, but no supernova or kilonova signature was detected. The burst properties and empirical correlations or distributions (e.g., duration, spectral hardness, location in the Amati correlation, $\varepsilon-$value, $f_{\rm eff}$ parameter, and physical offset) favor a compact binary merger origin. However, a dense circumburst medium with $n\sim 80\rm~cm^{-3}$, obtained by adopting the energy injection into a jet to interpret the late-time rebrightening is inconsistent with the compact binary merger origin. In this paper, we propose a two-component jet model to explain the multiwavelength afterglow observations of GRB 250221A, in which the relativistic narrow jet ($\rm \theta_{c} \sim 3.8^\circ$) produces the prompt and the early decay afterglow emission, while the mildly relativistic wide jet ($\rm \theta_{w} \sim 4.4^\circ$) dominates at later times, resulting in the observed rebrightening feature. If this is the case, one can obtain a lower medium density with $n\sim 0.72\rm~cm^{-3}$ which is a little bit higher than that of short GRBs in merger environments, but falls into the reasonable and acceptable range. Finally, a possible kilonova emission is also discussed within the scenario of compact star merger origin of GRB 250221A.
\end{abstract}


\section{Introduction}
Gamma-ray bursts (GRBs) are among the most energetic transients in the Universe. Observationally, based on the duration of prompt gamma-ray emission (e.g., $T_{\rm 90}$), GRBs are divided into two categories: long-duration GRBs (LGRBs) with $T_{\rm 90} > 2$ s, and short-duration GRBs (SGRBs) with $T_{\rm 90} < 2$ s \citep{1993ApJ...413L.101K, 2013ApJ...764..179B}. Several LGRBs have been confirmed to be associated with broad-lined Type Ic supernovae (SNe), and the progenitor of LGRBs are generally believed to be the collapse of massive stars \citep{1998Natur.395..670G, 2004ApJ...609L...5M, 2006ApJ...645L..21M, 2006Natur.442.1011P, 2006Natur.444.1010Z, 2012ApJ...759..107K, 2018ApJ...862..130L}. In contrast, mergers of compact binary star objects, such as neutron star-neutron star (NS–NS), or neutron star–black hole (NS–BH), are considered as possible progenitors of SGRBs \citep{2006Natur.444.1010Z, 2016NatCo...712898J, 2017ApJ...835..181L, 2019ApJ...870L..15L, 2019LRR....23....1M}. The “smoking gun” evidence is SGRB 170817A, in which both the kilonova (KN) electromagnetic emission AT2017gfo, powered by radioactive decay from r-process nuclei, and the corresponding gravitational-wave (GW) source GW170817 were simultaneously detected \citep{Abbott2017, 2019LRR....23....1M}. Whether it is the collapse of a massive star or the merger of compact stars, a hyperaccreting BH \citep{1999ApJ...518..356P, 2013ApJ...765..125L, 2017NewAR..79....1L}, or a rapidly spinning, strongly magnetized NS, named a millisecond magnetar \citep{1992Natur.357..472U, Dai1998a, Dai1998b, 2001ApJ...552L..35Z, 2011MNRAS.413.2031M, 2012MNRAS.419.1537B, 2014ApJ...785...74L} can be formed as the central engine of GRBs.  

The central engine drives an ultrarelativistic jet, which releases energy through internal shocks or magnetic reconnection to produce prompt gamma-ray emission \citep{1993ApJ...405..278M, 1994ApJ...430L..93R}, while the interaction of the jet with the external medium generates multiwavelength afterglow emission \citep{1997ApJ...476..232M,1998ApJ...497L..17S}. The structure of ultrarelativistic jets was traditionally considered as a top-hat, in which the energy density and Lorentz factor are constant within a core opening angle \citep{1997ApJ...487L...1R, 1999ApJ...519L..17S}. However, results of numerical simulations and observations of SGRB 170817A reveal that jets often possess an angular structure, such as a Gaussian distribution \citep{Zhang2002a, 2003ApJ...591.1075K, 2004ApJ...601L.119Z}, a power-law distribution \citep{1998ApJ...499..301M, 2002MNRAS.332..945R}, or two-component jets \citep{2004ApJ...605..300H, 2005ApJ...626..966P, 2008Natur.455..183R, 2011A&A...526A.113F}. In addition, when an ultrarelativistic jet crosses the stellar envelope or the merger ejecta, it can form a jet–cocoon system \citep{2017ApJ...834...28N, 2019ApJ...881...89L, 2018MNRAS.473..576G}, whose low-energy, wide-angle emission is particularly prominent in off-axis observations. 

From the observational point of view, the behavior of rebrightening in X-ray and optical afterglows has been detected with peak times typically occurring between $10^{\rm 2}$ and $10^{\rm 5}$ seconds \citep{2013ApJ...774...13L, 2024MNRAS.535.2482Y}. The physical interpretation of such a rebrightening feature is to invoke a structured jet with off-axis \citep{2003JCAP...10..005N, 2003Natur.426..154B, 2004ApJ...605..300H}, or late energy injection from slower ejecta or central engine activity (refreshed shock) \citep{Zhang2002b, 2004ApJ...615L..77B, 2015ApJ...814....1L}, or a medium density jump \citep{2002ApJ...565L..87D, 2002A&A...396L...5L, 2003ApJ...591L..21D}, or the existence of two-component jets \citep{2004ApJ...605..300H,2005MNRAS.357.1197W,2008Natur.455..183R}.

Recently, a short GRB 250221A, with a measured redshift of $z=0.768$ was detected, and the late-time afterglow of this GRB exhibited an unusual rebrightening feature $\sim$0.6 days postburst. The burst properties (duration, spectral hardness, peak energy, and location in the Amati relation) all favor a compact binary merger origin \citep{2026MNRAS.tmp..179A}, and suggest an internal energy injection from a long-lasting refreshed shock into the jet. However, it is found that the medium density of the circumburst is as high as $n \sim 80 \rm~cm^{-3}$, which is inconsistent with the typical density of the environment of merger systems \citep{2018pgrb.book.....Z}. So, \cite{2026MNRAS.tmp..179A} claimed that the GRB 250221A is likely originated from the scenario of massive star collapse. In this paper, by comparing the properties of GRB 250221A (e.g., $\varepsilon$ value, $f_{\rm eff}$ parameter, physical offset, and star formation rate (SFR)) with those of other typical LGRBs and SGRBs, we propose a two-component jet model to explain the multiwavelength afterglow observations of GRB 250221A. It is found that GRB 250221A is consistent with the scenario of compact star mergers.

The paper is organized as follows. The observations of the prompt and afterglow emission of the SGRB 250221A are presented in Section 2. In Section 3, we compare the properties of GRB 250221A with those of other long and short GRBs. The physical interpretation with a two-component jet model and the results are shown in Section 4. In Section 5, we attempt to calculate the possible kilonova emission to compare with observations. The conclusions are drawn in Section 6 with some additional discussion. Throughout this paper, we adopt a standard concordance cosmology with $\Omega_{\rm M} = 0.31$, $\Omega_{\rm \Lambda} = 0.69$, and $H_{0} = 67.7 ~\rm km ~\rm s^{-1} ~Mpc^{-1}$.

\section{The Observations and Data Analysis}
At 03:34:37 UT on 2025 February 21 (as $T_0$), the Burst Alert Telescope (BAT) of Swift triggered and located GRB 250221A, which was simultaneously detected by Konus-Wind \citep{2025GCN.39471....1P, 2025GCN.39423....1F}. The light curve of the prompt emission of this burst exhibits a single short-pulse structure with a duration of $T_{\rm 90} = 1.8 \pm 0.3$ s in 15-350 keV \citep{2025GCN.39471....1P}, and the prompt gamma-ray spectrum is well fitted by a power-law model with a photon index of $\Gamma = 1.43 \pm 0.23$ due to its narrow energy band. However, over the wide energy band range of Konus-Wind, the time-integrated spectrum of the burst can be fitted by a power-law with exponential cutoff (CPL) model, with a low-energy index of $\alpha = -1.05\pm0.12$ and a peak energy of $E_{\rm p}=242^{+26}_{-22}$ keV \citep{2025GCN.39423....1F}, and the fluence in the 10 keV -10 MeV band is $(1.67 \pm 0.09) \times 10^{-6}\rm~ erg~cm^{-2}$.

The Swift X-ray Telescope (XRT) began to observe the field from 107 s to 66.9 ks after the BAT trigger. We downloaded the public XRT data from the Swift archive \footnote{https://www.swift.ac.uk/xrt\_curves/01290305/}, and the X-ray light curve in 0.3–10 keV is well described by a power-law decay \citep{2025GCN.39414....1S}. Moreover, the Einstein Probe (EP), equipped with the Follow-up X-ray Telescope (FXT), conducted a follow-up observation of GRB 250221A approximately 2.1 days after the Swift/BAT trigger \citep{2025GCN.39491....1T}. We also collected the data of FXT from \cite{2026MNRAS.tmp..179A}.

Several optical telescopes reported the follow-up observations of GRB 250221A, such as the Very Large Telescope (VLT; \citealt{2025GCN.39418....1P}), the DDRAGO wide-field imager on the COLIBRI telescope \citep{2025GCN.39397....1W}, as well as radio data from the Australia Telescope Compact Array (ATCA; \citealt{2025GCN.39501....1G}) and the Very Large Array (VLA; \citealt{2025GCN.39433....1R}). We collected the optical data from \cite{2026MNRAS.tmp..179A}, and the optical light curve of GRB 250221A shows an initial decay followed by a rebrightening at around 0.6 days. Moreover, \cite{2026MNRAS.tmp..179A} analyzed the early afterglow spectrum of GRB 250221A, in which multiple emission and absorption lines were detected. They reported a redshift of $z=0.768$ for the host galaxy and confirmed that the GRB is located within this galaxy, with a projected offset of $<7^{+12}_{-7}$ kpc. They also derived an SFR of approximately $3~M_{\odot}~\rm yr^{-1}$ for the galaxy.

\section{Comparison of the Properties of GRB 250221A with Those of Other Long and Short GRBs}
Even the duration of a GRB can provide us with information on the timescale of central engine activity, but the duration is sensitively dependent on the observed energy and the instrument \citep{2013ApJ...763...15Q}. It suggests that the duration is no longer a reliable indicator of the physical origin of a GRB \citep{2009ApJ...703.1696Z, 2010ApJ...725.1965L}. For example, several LGRBs, i.e., GRB 060614 \citep{2006Natur.444.1044G, 2015NatCo...6.7323Y}, GRB 211211A \citep{2022Natur.612..223R, 2022Natur.612..228T, 2022Natur.612..232Y, 2023ApJ...943..146C, 2023NatAs...7...67G, 2025ApJ...988L..46L}, GRB 211227A \citep{2022ApJ...931L..23L, 2023A&A...678A.142F}, and GRB 230307A  \citep{2023ApJ...954L..29D, 2024Natur.626..737L, 2024Natur.626..742Y, 2024ApJ...962L..27D}, are thought to be from the merger of two compact stars. On the contrary, the SGRB 200826A is believed to originate from the collapse of massive stars \citep{2021NatAs...5..917A, 2021NatAs...5..911Z, 2022ApJ...932....1R}. So, \cite{2006Natur.444.1010Z} proposed adopting Type I (e.g., compact star mergers) and Type II (e.g., death of massive stars) to classify GRBs. If possible, the presence of supernova or kilonova emission associated with GRBs can directly provide evidence to determine the nature of the GRB progenitor \citep{2018pgrb.book.....Z}. However, catching such associated events remains challenging, especially the kilonova emission. From the statistical point of view, different progenitors of GRBs may follow different empirical correlations or distributions (e.g., Amati relation, $\varepsilon$ value, $f_{\rm eff}$ parameter, physical offset, and SFR). In this section, we compare several properties of GRB 250221A with those of other long and short GRBs.

\subsection{$E_{\rm p}-E_{\rm \gamma,iso}$ correlation}
Statistically, \cite{2002A&A...390...81A} discovered a relation between the peak energy (e.g., $E_{\rm p}$) and the isotropic-equivalent energy more energetic (e.g., $E_{\rm \gamma,iso}$) for most LGRBs, and it can be expressed as $E_{\rm p}(1+z) \propto E^{1/2}_{\rm \gamma,iso}$ (called the Amati relation). However, most Type I GRBs are inconsistent with Type II GRBs, and it seems to be a little shallower for the power index compared with that of Type II GRBs \citep{2009ApJ...703.1696Z}. Therefore, the Amati relation may play an important role in distinguishing between Type II and Type I GRBs. By adopting the values of $E_{\rm p}=242^{+26}_{-22}$ and $E_{\rm \gamma,iso}=(2.8 \pm 0.2) \times 10^{51}\rm~ erg$ from Konus-Wind observations \citep{2025GCN.39423....1F}, we plot GRB 250221A on the Amati relation diagram. It is found that GRB 250221A lies outside the $95\%$ confidence region of Type II GRBs, but seems to be closer to the Type I GRB population (see Figure 1(a)). This indicates that GRB 250221A is more likely to originate from a compact object merger (e.g., NS–NS or NS–BH) rather than from the collapse of a massive star.

\subsection{$\varepsilon$ value}
\cite{2010ApJ...725.1965L} proposed a new GRB classification method by invoking the values of $E_{\rm p}$ and $E_{\rm \gamma,iso}$, and introduced a parameter $\varepsilon$ which is defined as 
\begin{equation}\label{eq:1}
\varepsilon = E_{\rm \gamma,iso,52}/E^{\rm 5/3}_{\rm p,z,2},
\end{equation}
where $E_{\rm \gamma,iso,52}$ and $E_{\rm p,z,2}$ are expressed in units of $10^{52}\rm~ erg$ and $10^{2}\rm~keV$, respectively. Statistical analysis of GRB samples shows that $\varepsilon$ exhibits a bimodal distribution with $\varepsilon=0.03$ as the dividing line between the high- and low-$\varepsilon$ regions. In this classification method, GRBs located in the high-$\varepsilon$ (e.g., Type II) and low-$\varepsilon$ (e.g., Type I) regions are likely to be associated with the collapse of massive stars and compact star mergers, respectively. By adopting the method, one can obtain a value of $\varepsilon \sim 0.024$ for GRB 250221A, as shown in Figure 1(b). It is found that GRB 250221A is more likely to be located in the low-$\varepsilon$ region, but it is very close to the boundary between the long and short GRB populations. Therefore, GRB 250221A seems to be compatible with a compact object merger origin, but its ambiguous $\varepsilon$ value cannot provide strong evidence to support such an origin from the merger of compact objects.

\subsection{“Tip of the iceberg” effect}
The duration of GRBs may not be intrinsic to reflect the activity of the GRB central engine. For example, the majority of a GRB emission may fall below the background noise, and only the brightest pulses can be detected. As a result, the event can be observationally misclassified as short GRBs, a phenomenon known as the “tip of the iceberg” effect. In order to test this effect, \cite{2014MNRAS.442.1922L} introduced an amplitude parameter as a third dimension. The amplitude parameter is defined as $f_{\rm eff} = F^{'}_{\rm p} / F_{\rm B}$, where $F^{'}_{\rm p}$ denotes the peak flux of a pseudo-GRB, representing an intrinsically identical GRB with a lower amplitude, and $F_{\rm B}$ is the background flux. They found that the values of $f_{\rm eff}$ for most short GRBs are statistically larger than those for long GRBs, and suggested that most short GRBs are intrinsically short without being the “tip of iceberg” of long GRBs \citep{2014MNRAS.442.1922L}. By extracting the prompt emission light curve of GRB 250221A from Swift/BAT, one can calculate the value of $f_{\rm eff}\sim 1.68$ for GRB 250221A to adopt the same method from \cite{2014MNRAS.442.1922L}. We find that the location of GRB 250221A is consistent with the typical distribution of short GRBs (see Figure 1(c)), and suggest that GRB 250221A is an intrinsically short GRB, rather than due to the “tip of the iceberg” effect of long GRBs.

\subsection{Physical offset and star-forming rate of the host galaxy}
The host galaxies of LGRBs are typically actively star-forming and metal-poor, with burst locations that are closely offset from their host centers \citep{2006Natur.441..463F}. This is consistent with the model in which LGRBs originate from the collapse of massive stars \citep{1999ApJ...524..262M}. On the contrary, SGRBs are generally found in older stellar populations, such as elliptical galaxies with low SFRs, while some occur in star forming galaxies \citep{2005Natur.438..988B, 2014ARA&A..52...43B}. Their explosion sites are often offset from the centers of their host galaxies, with physical offsets ranging from a few to several tens of kiloparsecs \citep{2013ApJ...776...18F}. The Kolmogorov–Smirnov test shows that the physical offset distributions of long and short GRBs do not originate from the same parent population \citep{2016ApJS..227....7L}. We compare the physical offset of the host galaxy for GRB 250221A (e.g., $<7^{+12}_{-7}$ kpc) with that of other long and short GRB samples observed by the Hubble Space Telescope (see Figure 2(a)) \citep{2010ApJ...708....9F, 2016ApJ...817..144B}. It is found that the physical offset of GRB 250221A is larger than that of most LGRBs, but is statistically consistent with that of typical SGRBs. Moreover, we also compare the SFR (SFR $\sim 3~M_{\odot}~\rm yr^{-1}$) of the host galaxy for GRB 250221A with those of long \citep{2025ApJ...993...20D} and short \citep{2016ApJS..227....7L, 2022ApJ...940...57N} GRBs \footnote{The data of several short GRBs are taken from the  GHostS websit: https://www.grbhosts.org/} (see Figure 2(b)). We find that SFR alone cannot unambiguously determine the origin of GRB 250221A.

\section{Physical Interpretation}
The burst properties and empirical correlations or distributions (duration, spectral hardness, location in the Amati correlation, $\varepsilon$ value, $f_{\rm eff}$ parameter, and physical offset) are in favor of a compact binary merger origin (also see \cite{2026MNRAS.tmp..179A}). However, \cite{2026MNRAS.tmp..179A} adopt energy injection into a jet to interpret the late-time rebrightening in both X-ray and optical bands $\sim0.6$ days after the burst trigger, and obtain a dense circumburst medium with $n \sim 80\rm~cm^{-3}$. Such a high medium density is inconsistent with the environment of a compact binary merger, but is in favor of the typical environment of a collapsar \citep{2026MNRAS.tmp..179A}.

In order to reconcile the above contradictions, we invoke a two-component jet model, which was first proposed by \cite{2007ApJ...656L..57J} within the scenario of a compact binary merger to explain the multiwavelength afterglow of GRB 250221A. The two-component jet consists of an ultrarelativistic narrow jet core surrounded by a mildly relativistic wide jet component, and such a model is widely used to explain the complex evolution of GRB afterglows \citep{2003Natur.426..154B, 2005ApJ...626..966P, 2006MNRAS.370.1946G, 2008Natur.455..183R}. Within this scenario, the narrow jet is primarily responsible for the prompt gamma-ray and the early afterglow emission, whereas the emission from the wide, slow jet component rises to the late-time rebrightening afterglow. 

We employ the $VegasAfterglow$ code, which is proposed by \cite{2026JHEAp..5000490W} to fit the multiwavelength afterglow data of GRB 250221A. This code describes an axisymmetric jet model, and the two-component jet structure can be expressed as follows: 
\begin{equation}\label{eq:1}
\Gamma_0(\theta) =
\begin{cases}
\Gamma_0, & 0 < \theta \le \theta_c\\[6pt]
\Gamma_{\rm 0,w}, & \theta_c < \theta \le \theta_w\\[6pt]
1, & \theta_w < \theta
\end{cases}\,,
\end{equation}

\begin{equation}\label{eq:2}
\dfrac{dE}{d\Omega}(\theta) =
\begin{cases}
\dfrac{E_{\rm K,iso}}{4\pi}, & 0<\theta \le \theta_c \\[6pt]
\dfrac{E_{\rm K,iso,w}}{4\pi}, & \theta_c < \theta \le \theta_w\\[6pt]
0, & \theta_w < \theta
\end{cases}\,.
\end{equation}
Here, $E_{\rm K,iso}$, $\rm \Gamma_{0}$ and $\rm \theta_{c}$ represent the isotropic equivalent kinetic energy, initial bulk Lorentz factor, and half-opening angle of the narrow jet core, respectively. $E_{\rm K,iso,w}$, $\rm \Gamma_{0,w}$ and $\rm \theta_{w}$ correspond to the parameters of the wide jet component. By assuming a constant density profile of the ISM for the afterglow, we adopt the Markov Chain Monte Carlo (MCMC) method provided by \textit{Vegasafterglow} \citep{2026JHEAp..5000490W} to perform the posterior parameter inference\footnote{ Moreover, several other public codes of afterglow are also proposed in the literature to fit the afterglow data, such as \textit{afterglowpy} \citep{2020ApJ...896..166R,2024ApJ...975..131R}, \textit{jetsimpy} \citep{2024ApJS..273...17W,2025ApJ...990..110W}, \textit{ASGARD} \citep{2024ApJ...962..115R}, and \textit{redback} \citep{2024MNRAS.531.1203S}.}. The basic parameters, $E_{\rm K,iso}$, $\rm \Gamma_{0}$, $\rm \theta_{c}$, $E_{\rm K,iso,w}$, $\rm \Gamma_{0,w}$, $\rm \theta_{w}$, $n$, $p$, $\epsilon_{\rm e}$, and $\epsilon_{\rm B}$, are shown in Table 1. Of these, the parameters $\rm \theta_{c}$, $\rm \theta_{w}$, and $p$ are sampled on a linear scale, while the remaining parameters are sampled on a logarithmic scale. The fraction of electrons $\xi_{\rm e}=1$ is fixed as commonly adopted in afterglow modeling. The best-fit results for the multiwavelength afterglow light curves are shown in Figure 3, and the model parameters are summarized in Table 1. The corner plots of the posterior density distributions from the MCMC samples are shown in Figure 4.

The fitting results show that the afterglow of GRB 250221A can be explained by a narrow jet core with a half-opening angle of $\rm \theta_{c}\sim 3.8^\circ$ and an initial bulk Lorentz factor $\rm \Gamma_{0}\sim 507$, surrounded by a mildly relativistic wide jet with $\rm \theta_{w}\sim 4.4 ^\circ$ and $\rm \Gamma_{0,w} \sim 46$. Within this scenario, the emission from the narrow jet has entered its decay phase at the beginning of the observation of the afterglow, while the emission from the wide jet continues to rise and eventually becomes dominant. It can result in the distinct rebrightening at around 0.6 days observed in the optical and X-ray bands. Within the framework of the two-component jet model, the derived circumburst medium density is $n\sim 0.72~\rm cm^{-3}$, which is much lower than that of previous work by adopting an energy-injection scenario (e.g., \cite{2026MNRAS.tmp..179A}) inferred a density around $n\sim 80~\rm cm^{-3}$. Moreover, we compare the obtained medium density with that of other short GRBs taken from \cite{2015ApJ...815..102F} (see right of Figure 3), and find that although this medium density is slightly higher than that of SGRBs in merger environments, it still falls into a reasonable and acceptable range \citep{2015ApJ...815..102F}. Together with the burst properties and empirical correlations or distributions of GRB 250221A, it suggests that the SGRB 250221A originated from the merger of compact stars.

\section{Possible kilonova emission}
If GRB 250221A is indeed originated from the mergers of compact stars, an optical/infrared transient (called a kilonova) can be powered by the radioactive decay of r-process nuclei \citep{1998ApJ...507L..59L, 2010MNRAS.406.2650M, 2011ApJ...732L...6R, 2013PhRvD..87b4001H, 2013MNRAS.430.1061R, 2021ApJ...912...14Y, 2025ApJ...988L..46L}. During the merger process, a large amount of neutron-rich matter can be ejected through tidal tails and disk winds. The r-process nucleosynthesis within the ejecta produces unstable heavy nuclei whose radioactive decay heats the ejecta and provides the energy for kilonova emission \citep{1998ApJ...507L..59L, 2010MNRAS.406.2650M, 2013ApJ...776L..40Y, 2017ApJ...837...50G,2020NatAs...4...77J}. In this section, we attempt to calculate the possible kilonova emission at $z=0.768$ to compare with the observations.

Follow-up optical observations of GRB 250221A from the COLIBRI and VLT reveal a light curve showing an initial decay phase and a rebrightening at approximately 0.6 days \citep{2026MNRAS.tmp..179A}. In order to investigate whether this feature could be caused by the possible associated kilonova emission, we calculate the possible kilonova emission in the $i$ and $r$ bands by adopting $M_{\rm ej}=(10^{-3}-10^{-1})\rm~ M_{\odot}$, velocity $\beta = 0.1$ and opacity $\kappa = 1 ~\rm cm^{2}~g^{-1}$ \citep{2013PhRvD..87b4001H}. In our calculation, we only consider the r-process itself, and the equations to describe the evolution of the ejecta dynamics and the internal energy are taken from \citep{2013ApJ...776L..40Y} and \citep{2021ApJ...912...14Y}. The results of the numerical calculations and the comparison with observations are shown in Figure 5. We find that the potential kilonova emission associated with GRB 250221A at a redshift of $z = 0.768$ would be extremely faint. Even at its peak, the kilonova flux remains 2 orders of magnitude lower than the observed optical data. Therefore, the rebrightening feature observed in the optical afterglow of GRB 250221A is unlikely to be caused by kilonova emission.

On the other hand, one possibility is that the central engine is a magnetar, which may be formed after the mergers of binary neutron stars, and the spin energy of magnetar can be injected into kilonova to increase its peak luminosity \citep{2013ApJ...776L..40Y,2014MNRAS.439.3916M,2017ApJ...837...50G,2024MNRAS.527.5166W,2025ApJ...978...52A}. If this is the case, it requires an extreme spin-down luminosity and timescale of a magnetar to power such optical rebrightening for GRB 250221A. However, there is a lack of enough observational evidence to support that the magnetar resides as the central engine of GRB 250221A.

\section{Conclusion and Discussion}
GRB 250221A is a short-duration burst simultaneously detected by Swift and Konus-Wind with a duration of $T_{90} \sim 1.8$ s at a redshift of $z = 0.768$. The prompt gamma-ray emission of this burst consists of a single short pulse, and the CPL model can well fit the spectrum of GRB 250221A with a peak energy $E_{\rm p}=242^{+26}_{-22}$ keV. By adopting the measured redshift $z = 0.768$ \citep{2026MNRAS.tmp..179A}, one can estimate the isotropic gamma-ray energy $E_{\rm \gamma,iso} = (2.8 \pm 0.2) \times 10^{51}\rm~ erg$ \citep{2025GCN.39423....1F}. The physical offset and SFR of the host galaxy of GRB 250221A are about $< 7^{+12}_{-7}$ kpc and $\sim 3~M_{\odot}~\rm yr^{-1}$, respectively. More interestingly, both the optical and X-ray afterglow of GRB 250221A exhibit an initial decay followed by a remarkable rebrightening around 0.6 days after the trigger.

By comparing the properties and empirical correlations or distributions (e.g., duration, spectral hardness, location in the Amati correlation, $\varepsilon$ value, $f_{\rm eff}$ parameter, and physical offset) of GRB 250221A with those of other Type I and Type II GRBs, it is found that GRB 250221A shares similar properties and empirical correlations or distributions with Type I GRBs, suggesting that GRB 250221A favors a compact binary merger origin. However, \cite{2026MNRAS.tmp..179A} invoked a model of energy injection into a jet to interpret the late-time rebrightening, and obtained a medium density as high as $n \sim 80 \rm~cm^{-3}$ , which is inconsistent with typical merger environments \citep{2015ApJ...815..102F}, but is in favor of a collapsar environment.

Alternatively, in order to reconcile the above contradictions, we propose a two-component jet model within the scenario of compact binary merger to explain the multiwavelength afterglow of GRB 250221A. The narrow jet is primarily responsible for the prompt gamma-ray emission and the early afterglow emission, whereas the emission from the wide, slow jet component rises to the late-time rebrightening afterglow. Within this scenario, one can obtain the relativistic narrow jet ($\rm \theta_{c} \sim 3.8^\circ$, $\rm \Gamma_{0} \sim 507$) which produces an initial rapid decay phase after deceleration, and the mildly relativistic wide jet ($\rm \theta_{w} \sim 4.4^\circ$, $\rm \Gamma_{0,w} \sim 46$) dominates the late-time emission to result in a pronounced rise in the light curve around 0.6 days after the trigger. On the other hand, one can obtain the medium density $n\sim 0.72~\rm cm^{-3}$, which is much lower than the value calculated in \cite{2026MNRAS.tmp..179A} by invoking the energy injection into a jet. Although such medium density is slightly higher than that of SGRBs in merger environments, it still falls within the reasonable and acceptable range \citep{2015ApJ...815..102F}. In any case, together with the burst properties and empirical correlations or distributions of GRB 250221A, as well as the lower medium density, it suggests that the SGRB 250221A still originates from the mergers of compact stars.

The fitting results by adopting the two-component jet model are dependent on the selected structure of the jet. For example, \cite{2024ApJ...962..115R} performed a top-hat inner core encompassed by a power-law-structured outer wing as the structure of the two-component jet. Moreover, the algorithm of different afterglow codes (e.g., afterglowpy, jetsimpy, ASGARD, and redback) can also result in a little difference even for the same data. On the other hand, one needs to clarify that the two-component jet model we adopt is not the only physical model to interpret the observational afterglow data of GRB 250221A, and we cannot rule out other possible physical mechanisms \citep{2024MNRAS.535.2482Y}. However, together with the burst properties and empirical correlations or distributions of GRB 250221A, it provides at least a self-consistent physical explanation of the data. 

Within the scenario of the two-component jet model, the problem is that the wide component possesses higher energy than the narrow core, while their half-opening angles differ by only about 0.01 rad. Based on the 3-D simulations from \cite{2021MNRAS.502.1843N}, they adopted a continuous angular structure in which both the energy and the Lorentz factor peak at finite angles away from the axis, often surrounded by slower, baryon-rich outflows, and it is invoked to interpret off-axis GRB and its kilonova emission. The two‑component model is fundamentally a parameterized approximation designed to capture strong angular stratification within the jet, rather than to represent two physically distinct jet layers. However, we adopt the two-component picture of a narrow-fast core encircled by a wide-slower component, and it is quite different from the picture in \cite{2021MNRAS.502.1843N}.

Moreover, we also calculate the theoretical kilonova emission that is possibly associated with GRB 250221A in the $i-$ and $r$ bands. It is found that the potential kilonova emission associated with GRB 250221A at a redshift of $z = 0.768$ would be extremely faint. Even at its peak, the kilonova emission remains two orders of magnitude lower than the observed optical data. Therefore, the rebrightening feature observed in the optical afterglow of GRB 250221A is unlikely to be caused by the kilonova emission.

\section{Acknowledgments}
We acknowledge the use of the public data from the Swift/BAT and XRT Science Data Center. This work is supported by the Natural Science Foundation of China (grant Nos. 12494574 and 12494570), the Natural Science Foundation of Guangxi (grant Nos. 2023GXNSFDA026007 and 2025GXNSFDA02850010), the Program of Bagui Scholars Program (LHJ), and the Guangxi Talent Program (Highland of Innovation Talents).

{}

\clearpage
\begin{table*}[h]\footnotesize %
 \centering
  \begin{threeparttable}
  \caption{Best-fitting results with two-component jet model}
  \setlength{\tabcolsep}{2mm}{
  \begin{center}  
  \renewcommand\arraystretch{1.5}
  \begin{tabular}{ccccc}
  \hline    
  \hline
  $\rm Parameter$   & $\rm Scale^a$ & $\rm Range$ & $\rm Value$  \\ 
  \hline
  $E_{\rm K,iso} (\rm erg)$ & LOG & [$10^{52}, 10^{55}$] & $\rm 52.689^{+0.121}_{-0.106}$ \\
  $\rm \Gamma_{0} $ & LOG & [100, 800] & $\rm 2.705^{+0.112}_{-0.109}$ \\
  $\rm \theta_{c} (\rm rad)$ & LINEAR & [0, 0.1] & $\rm 0.067^{+0.004}_{-0.003}$ \\
  $E_{\rm K,iso,w} (\rm erg)$ & LOG & [$10^{52}, 5\times10^{54}$]  & $\rm 53.954^{+0.141}_{-0.187}$ \\
  $\rm \Gamma_{0,w} $ & LOG & [10, 100] & $\rm 1.660^{+0.155}_{-0.139}$ \\
  $\rm \theta_{w} (\rm rad)$ & LINEAR &  [0.07, 0.25] &  $\rm 0.076^{+0.006}_{-0.004}$ \\
  $n$ ($\rm cm^{-3}$)& LOG & [$10^{-3}$, 1] &  $\rm -0.142^{+0.090}_{-0.114}$ \\
  $p$ & LINEAR &  [2, 3] &  $\rm 2.004^{+0.001}_{-0.001}$ \\
  $\epsilon_{\rm e}$ & LOG &  [$10^{-3}, 0.5$] &  $\rm -0.505^{+0.060}_{-0.055}$ \\
  $\epsilon_{\rm B}$ & LOG & [$10^{-3}, 0.1$]  & $\rm -1.692^{+0.090}_{-0.086}$ \\
  \hline
\end{tabular}
\begin{tablenotes}
\footnotesize
\item $^a$ LOG means we sample $\rm log_{10}$(x), LINEAR means we sample linearly.
\end{tablenotes}
\end{center}}
\end{threeparttable}
\end{table*}

\begin{figure*}[htbp!]\label{fig:1}
\center
\includegraphics[angle=0,width=0.45\textwidth]{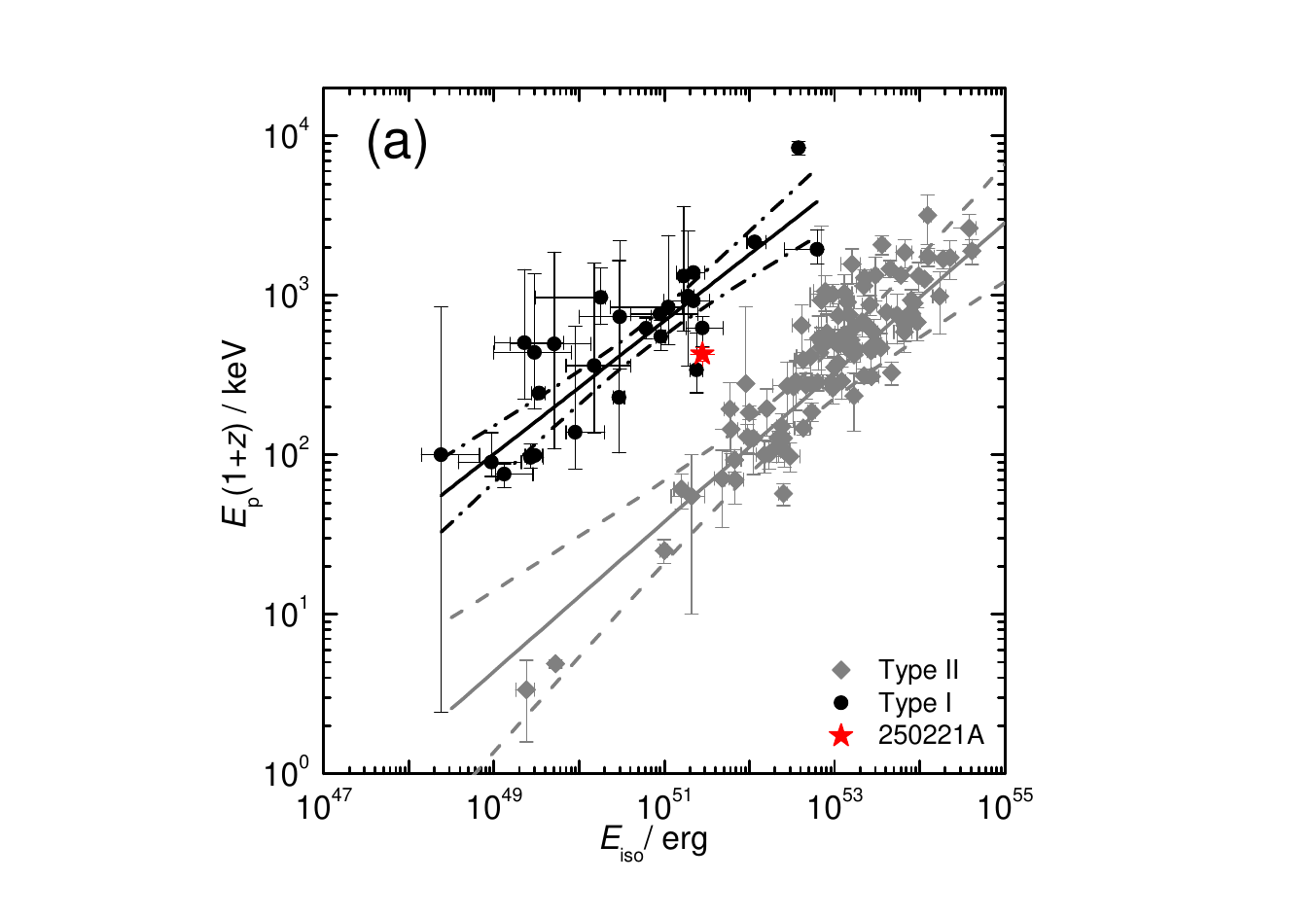}
\includegraphics[angle=0,width=0.5\textwidth]{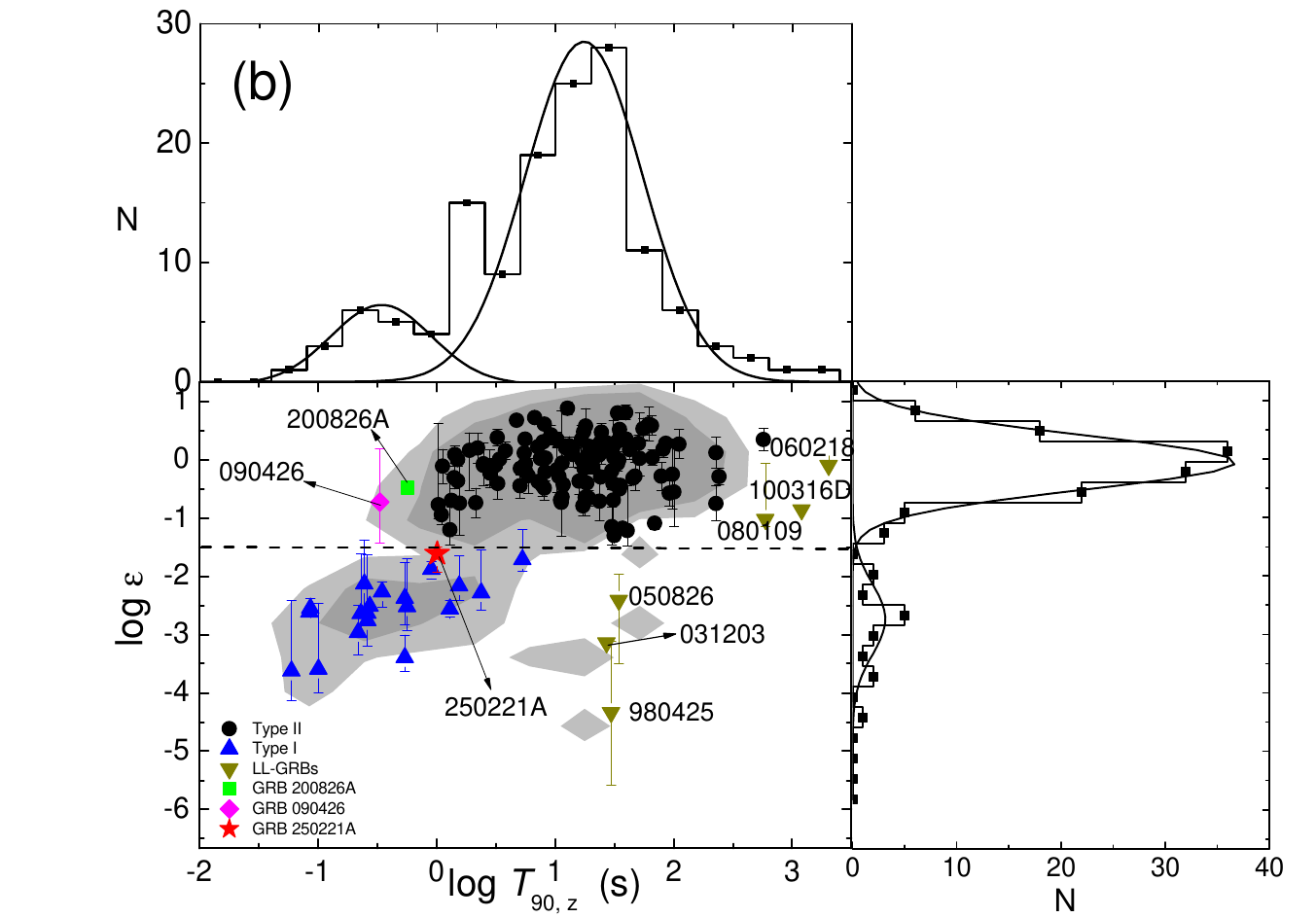}
\includegraphics[angle=0,width=0.5\textwidth]{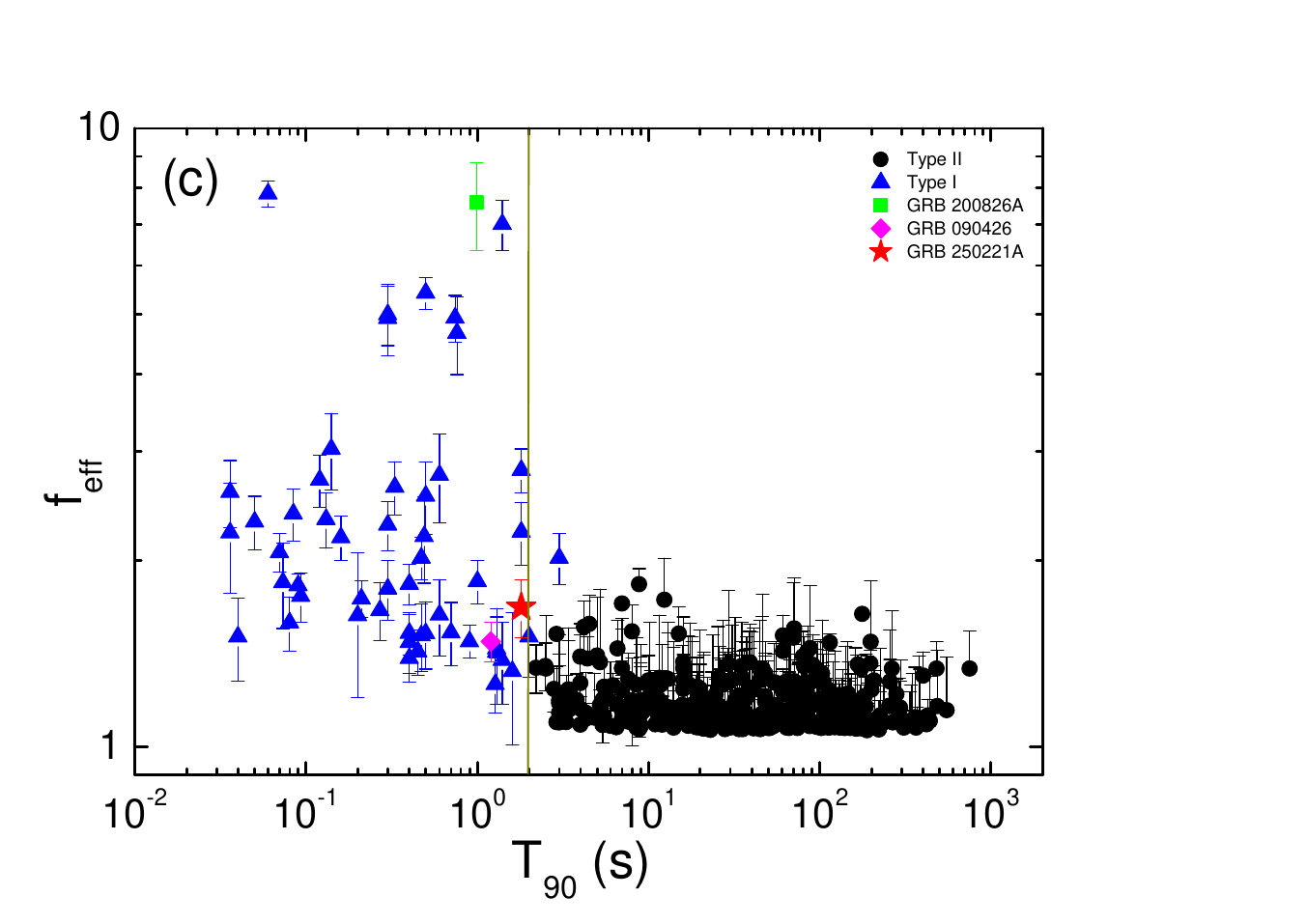}
\caption{(a) $E_{\rm p}-E_{\rm \gamma,iso}$ correlation plane. Gray diamonds and black-filled circles represent Type II and Type I GRBs, respectively. (b) 1D and 2D distributions in the $T_{90}-\varepsilon$ plane. The dashed line marks the empirical boundary $\varepsilon = 0.03$. (c) The $f_{\rm eff}-T_{\rm 90}$ distribution. The vertical solid line denotes the traditional classification boundary at $T_{90}$ = 2 s. The data adopted above are taken from \cite{2002A&A...390...81A, 2009ApJ...703.1696Z, 2010ApJ...725.1965L, 2014MNRAS.442.1922L}.}
\end{figure*}
\begin{figure*}[htbp!]\label{fig:2}
\center
\includegraphics[angle=0,width=0.5\textwidth]{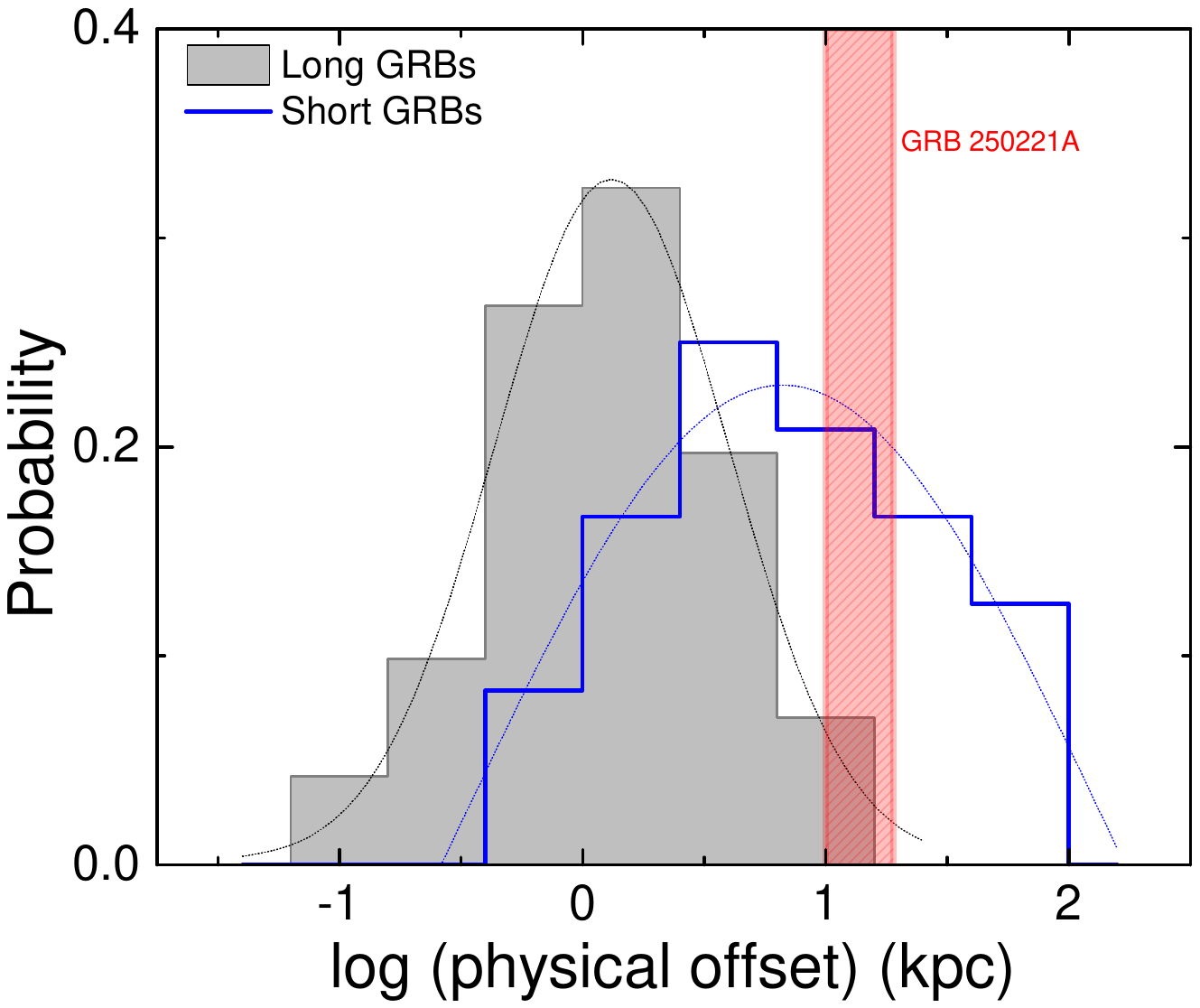}
\includegraphics[angle=0,width=0.45\textwidth]{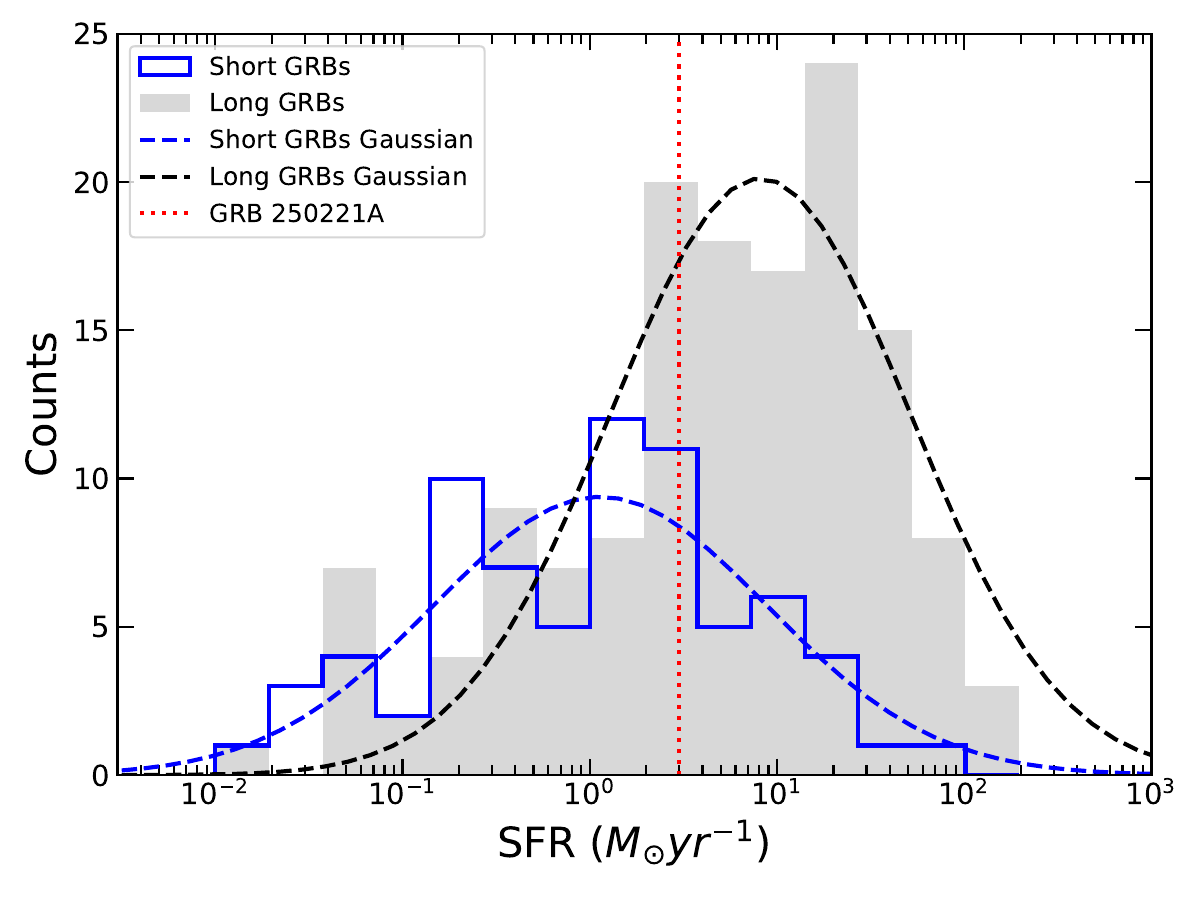}
\caption{Left: physical offset distributions of long (gray; \cite{2016ApJ...817..144B}) and short (blue; \cite{2010ApJ...708....9F}) GRBs. Right: distribution of star formation rates of host galaxies for long (gray; \cite{2025ApJ...993...20D}) and short (blue; \cite{2016ApJS..227....7L} and  \cite{2022ApJ...940...57N}) GRBs. The dotted line is the best Gaussian fit, and the red vertical line is GRB 250221A.}
\end{figure*}
\begin{figure*}[htbp!]\label{fig:3}
\center
\includegraphics[angle=0,width=0.45\textwidth]{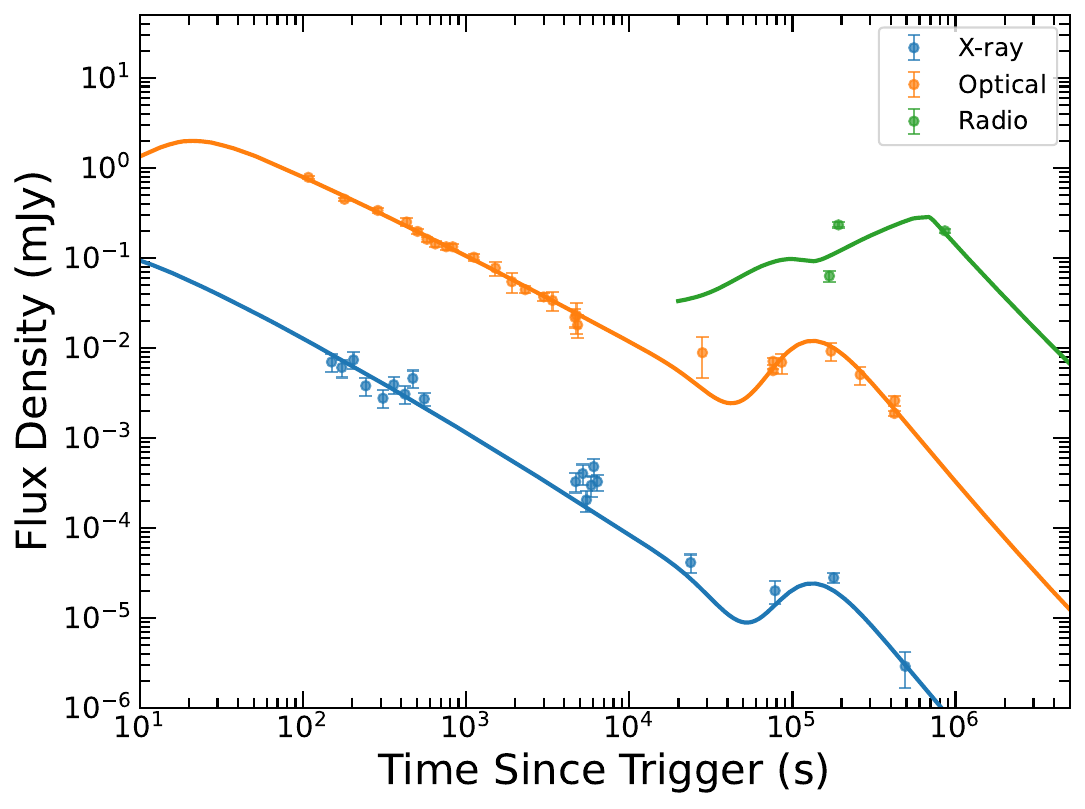}
\includegraphics[angle=0,width=0.45\textwidth]{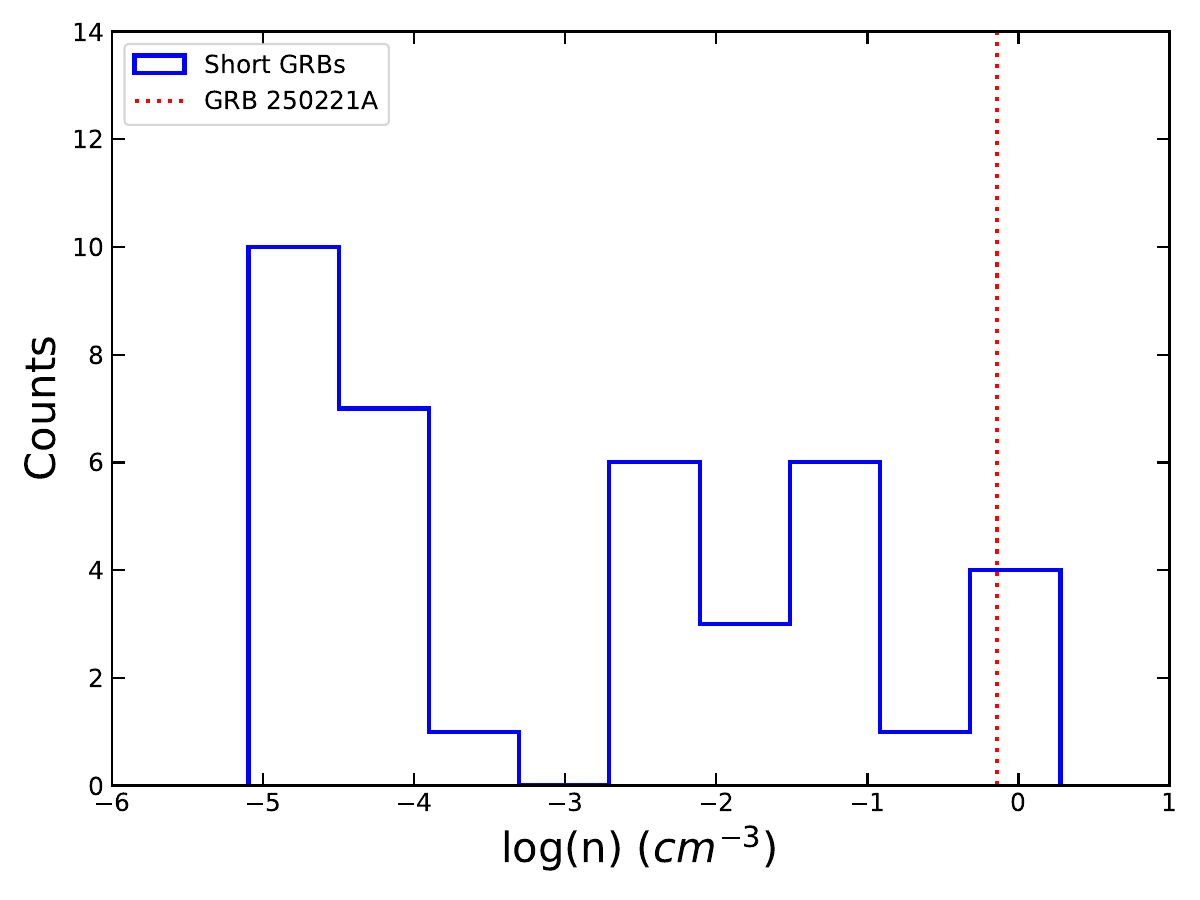}
\caption{Left: best-fit multiwavelength afterglow light curves with the two-component jet model (solid lines). The blue circles represent the X-ray data (at 1 keV), while the optical (orange) and radio (green) observational data are taken from \cite{2026MNRAS.tmp..179A}. Right: distribution of circumburst medium density for short GRBs \citep{2015ApJ...815..102F}. The red vertical dotted line is the inferred circumburst density (e.g., $n\sim 0.72~\rm cm^{-3}$) of GRB 250221A.}
\end{figure*}
\begin{figure*}[htbp!]\label{fig:4}
\center
\includegraphics[angle=0,width=1\textwidth]{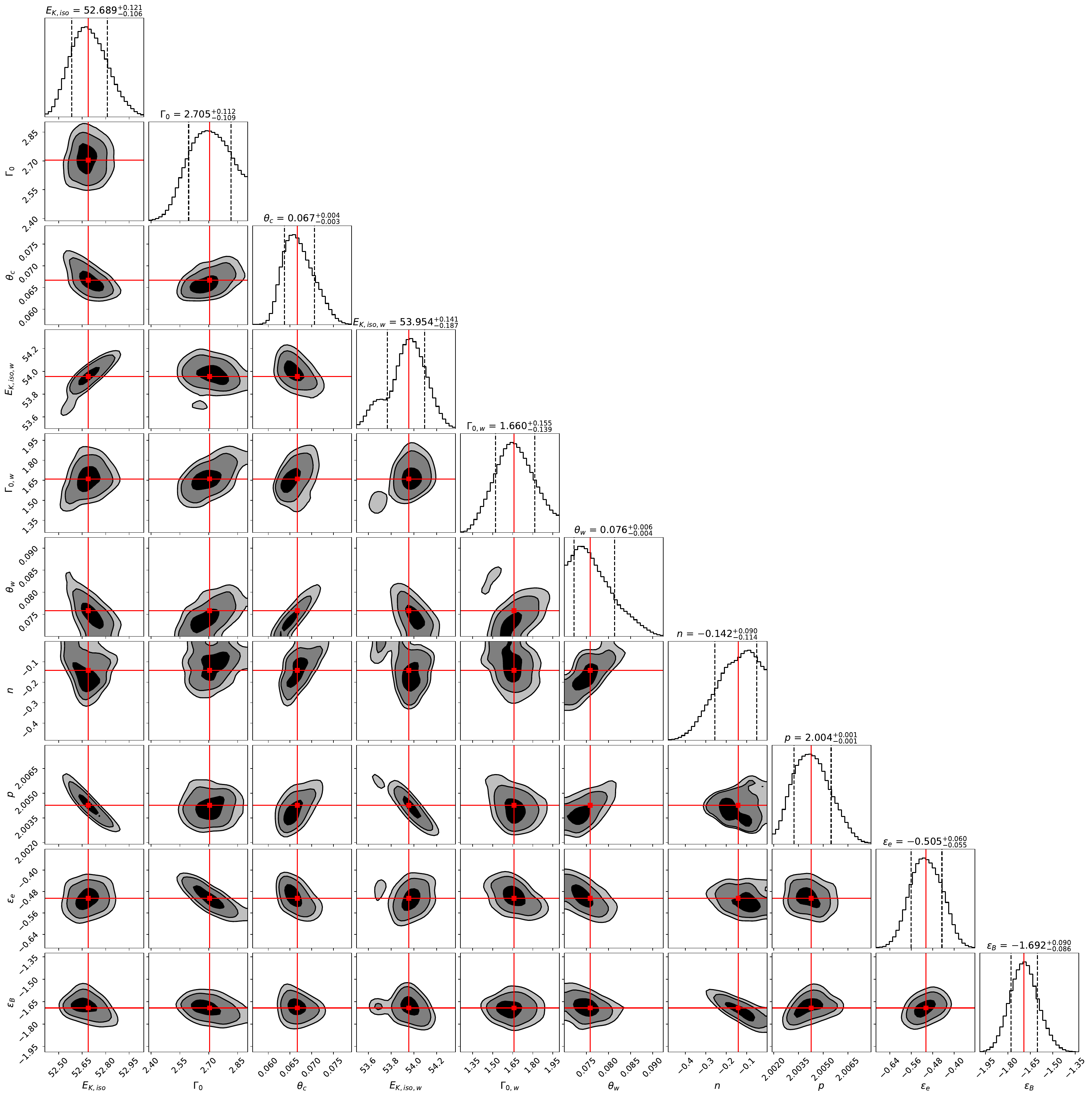}
\caption{Corner diagram of the posterior distributions with the MCMC method by adopting the two-component jet model.}
\end{figure*}
\begin{figure*}[htbp!]\label{fig:5}
\center
\includegraphics[angle=0,width=0.5\textwidth]{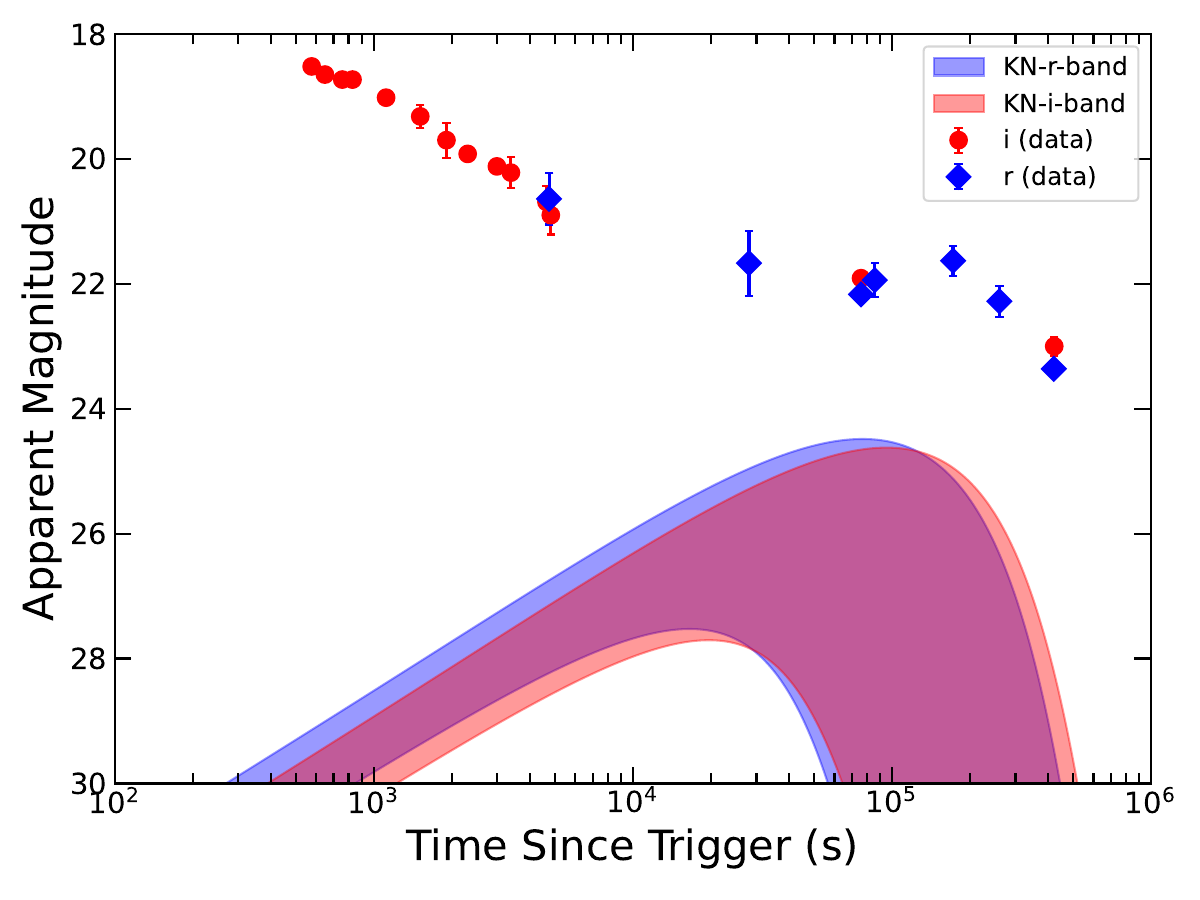}
\caption{Kilonova emission in the $i$ band (red) and the $r$ band (blue). The parameters are adopted as ejecta masses $M_{\rm ej} = 10^{-3}$--$10^{-1}~M_{\odot}$, ejecta velocity $\beta = 0.1$, and opacity $\kappa = 1~\rm cm^2~g^{-1}$. The red circles and blue diamonds are the observational data in the optical $i$ band and $r$ band, respectively.}
\end{figure*}


\begin{thebibliography}{}
\bibitem[Abbott et al.(2017)]{Abbott2017} Abbott, B.~P., Abbott, R., Abbott, T.~D., et al.\ 2017, \prl, 119, 161101. doi:10.1103/PhysRevLett.119.161101
\bibitem[Ahumada et al.(2021)]{2021NatAs...5..917A} Ahumada, T., Singer, L.~P., Anand, S., et al.\ 2021, Nature Astronomy, 5, 917. doi:10.1038/s41550-021-01428-7
\bibitem[Ai et al.(2025)]{2025ApJ...978...52A} Ai, S., Gao, H., \& Zhang, B.\ 2025, \apj, 978, 1, 52. doi:10.3847/1538-4357/ad93b4
\bibitem[Amati et al.(2002)]{2002A&A...390...81A} Amati, L., Frontera, F., Tavani, M., et al.\ 2002, \aap, 390, 81. doi:10.1051/0004-6361:20020722
\bibitem[Angulo-Valdez et al.(2026)]{2026MNRAS.tmp..179A} Angulo-Valdez, C., Becerra, R.~L., Gill, R., et al.\ 2026, \mnras. doi:10.1093/mnras/stag184, in press
\bibitem[Berger et al.(2003)]{2003Natur.426..154B} Berger, E., Kulkarni, S.~R., Pooley, G., et al.\ 2003, \nat, 426, 6963, 154. doi:10.1038/nature01998
\bibitem[Berger et al.(2005)]{2005Natur.438..988B} Berger, E., Price, P.~A., Cenko, S.~B., et al.\ 2005, \nat, 438, 7070, 988. doi:10.1038/nature04238
\bibitem[Berger(2014)]{2014ARA&A..52...43B} Berger, E.\ 2014, \araa, 52, 43. doi:10.1146/annurev-astro-081913-035926
\bibitem[Bj{\"o}rnsson et al.(2004)]{2004ApJ...615L..77B} Bj{\"o}rnsson, G., Gudmundsson, E.~H., \& J{\'o}hannesson, G.\ 2004, \apjl, 615, 2, L77. doi:10.1086/426477
\bibitem[Blanchard et al.(2016)]{2016ApJ...817..144B} Blanchard, P.~K., Berger, E., \& Fong, W.-. fai .\ 2016, \apj, 817, 2, 144. doi:10.3847/0004-637X/817/2/144
\bibitem[Bromberg et al.(2013)]{2013ApJ...764..179B} Bromberg, O., Nakar, E., Piran, T., et al.\ 2013, \apj, 764, 2, 179. doi:10.1088/0004-637X/764/2/179
\bibitem[Bucciantini et al.(2012)]{2012MNRAS.419.1537B} Bucciantini, N., Metzger, B.~D., Thompson, T.~A., et al.\ 2012, \mnras, 419, 2, 1537. doi:10.1111/j.1365-2966.2011.19810.x
\bibitem[Chang et al.(2023)]{2023ApJ...943..146C} Chang, X.-Z., L{\"u}, H.-J., Yang, X., et al.\ 2023, \apj, 943, 2, 146. doi:10.3847/1538-4357/aca969
\bibitem[Dai \& Lu(2002)]{2002ApJ...565L..87D} Dai, Z.~G. \& Lu, T.\ 2002, \apjl, 565, 2, L87. doi:10.1086/339418
\bibitem[Dai \& Wu(2003)]{2003ApJ...591L..21D} Dai, Z.~G. \& Wu, X.~F.\ 2003, \apjl, 591, 1, L21. doi:10.1086/377037
\bibitem[Dai \& Lu(1998a)]{Dai1998a} Dai, Z.~G., \& Lu, T.\ 1998a, \aap, 333, L87. doi:10.48550/arXiv.astro-ph/9810402
\bibitem[Dai \& Lu(1998b)]{Dai1998b} Dai, Z.~G., \& Lu, T.\ 1998b, \prl, 81, 4301. doi:10.1103/PhysRevLett.81.4301
\bibitem[Dichiara et al.(2023)]{2023ApJ...954L..29D} Dichiara, S., Tsang, D., Troja, E., et al.\ 2023, \apjl, 954, 1, L29. doi:10.3847/2041-8213/acf21d
\bibitem[Dong et al.(2025)]{2025ApJ...993...20D} Dong, X.-F., Huang, Y.-F., Zhang, Z.-B., et al.\ 2025, \apj, 993, 1, 20. doi:10.3847/1538-4357/ae04e7
\bibitem[Du et al.(2024)]{2024ApJ...962L..27D} Du, Z., L{\"u}, H., Yuan, Y., et al.\ 2024, \apjl, 962, 2, L27. doi:10.3847/2041-8213/ad22e2
\bibitem[Ferro et al.(2023)]{2023A&A...678A.142F} Ferro, M., Brivio, R., D'Avanzo, P., et al.\ 2023, \aap, 678, A142. doi:10.1051/0004-6361/202347113
\bibitem[Filgas et al.(2011)]{2011A&A...526A.113F} Filgas, R., Kr{\"u}hler, T., Greiner, J., et al.\ 2011, \aap, 526, A113. doi:10.1051/0004-6361/201015320
\bibitem[Fong et al.(2010)]{2010ApJ...708....9F} Fong, W., Berger, E., \& Fox, D.~B.\ 2010, \apj, 708, 1, 9. doi:10.1088/0004-637X/708/1/9
\bibitem[Fong \& Berger(2013)]{2013ApJ...776...18F} Fong, W. \& Berger, E.\ 2013, \apj, 776, 1, 18. doi:10.1088/0004-637X/776/1/18
\bibitem[Fong et al.(2015)]{2015ApJ...815..102F} Fong, W., Berger, E., Margutti, R., et al.\ 2015, \apj, 815, 2, 102. doi:10.1088/0004-637X/815/2/102
\bibitem[Frederiks et al.(2025)]{2025GCN.39423....1F} Frederiks, D., Lysenko, A., Ridnaya, A., et al.\ 2025, GCN, 39423, 1.
\bibitem[Fruchter et al.(2006)]{2006Natur.441..463F} Fruchter, A.~S., Levan, A.~J., Strolger, L., et al.\ 2006, \nat, 441, 7092, 463. doi:10.1038/nature04787
\bibitem[Galama et al.(1998)]{1998Natur.395..670G} Galama, T.~J., Vreeswijk, P.~M., van Paradijs, J., et al.\ 1998, \nat, 395, 6703, 670. doi:10.1038/27150
\bibitem[Gao et al.(2017)]{2017ApJ...837...50G} Gao, H., Zhang, B., L{\"u}, H.-J., et al.\ 2017, \apj, 837, 1, 50
\bibitem[Gehrels et al.(2006)]{2006Natur.444.1044G} Gehrels, N., Norris, J.~P., Barthelmy, S.~D., et al.\ 2006, \nat, 444, 7122, 1044. doi:10.1038/nature05376
\bibitem[Gompertz et al.(2023)]{2023NatAs...7...67G} Gompertz, B.~P., Ravasio, M.~E., Nicholl, M., et al.\ 2023, Nature Astronomy, 7, 67. doi:10.1038/s41550-022-01819-4
\bibitem[Gottlieb et al.(2018)]{2018MNRAS.473..576G} Gottlieb, O., Nakar, E., \& Piran, T.\ 2018, \mnras, 473, 1, 576. doi:10.1093/mnras/stx2357
\bibitem[Granot et al.(2006)]{2006MNRAS.370.1946G} Granot, J., K{\"o}nigl, A., \& Piran, T.\ 2006, \mnras, 370, 4, 1946. doi:10.1111/j.1365-2966.2006.10621.x
\bibitem[Gulati et al.(2025)]{2025GCN.39501....1G} Gulati, A., Anderson, G.~E., Morley, C., et al.\ 2025, GCN, 39501, 1.
\bibitem[Hotokezaka et al.(2013)]{2013PhRvD..87b4001H} Hotokezaka, K., Kiuchi, K., Kyutoku, K., et al.\ 2013, \prd, 87, 024001. doi:10.1103/PhysRevD.87.024001
\bibitem[Huang et al.(2004)]{2004ApJ...605..300H} Huang, Y.~F., Wu, X.~F., Dai, Z.~G., et al.\ 2004, \apj, 605, 1, 300. doi:10.1086/382202
\bibitem[Jin et al.(2007)]{2007ApJ...656L..57J} Jin, Z.~P., Yan, T., Fan, Y.~Z., et al.\ 2007, \apjl, 656, 2, L57. doi:10.1086/512971
\bibitem[Jin et al.(2016)]{2016NatCo...712898J} Jin, Z.-P., Hotokezaka, K., Li, X., et al.\ 2016, Nature Communications, 7, 12898. doi:10.1038/ncomms12898
\bibitem[Jin et al.(2020)]{2020NatAs...4...77J} Jin, Z.-P., Covino, S., Liao, N.-H., et al.\ 2020, Nature Astronomy, 4, 77. doi:10.1038/s41550-019-0892-y
\bibitem[Kelly \& Kirshner(2012)]{2012ApJ...759..107K} Kelly, P.~L. \& Kirshner, R.~P.\ 2012, \apj, 759, 2, 107. doi:10.1088/0004-637X/759/2/107
\bibitem[Kouveliotou et al.(1993)]{1993ApJ...413L.101K} Kouveliotou, C., Meegan, C.~A., Fishman, G.~J., et al.\ 1993, \apjl, 413, L101. doi:10.1086/186969
\bibitem[Kumar \& Granot(2003)]{2003ApJ...591.1075K} Kumar, P. \& Granot, J.\ 2003, \apj, 591, 2, 1075. doi:10.1086/375186
\bibitem[Lamb et al.(2019)]{2019ApJ...870L..15L} Lamb, G.~P., Lyman, J.~D., Levan, A.~J., et al.\ 2019, \apjl, 870, 2, L15. doi:10.3847/2041-8213/aaf96b
\bibitem[Laskar et al.(2015)]{2015ApJ...814....1L} Laskar, T., Berger, E., Margutti, R., et al.\ 2015, \apj, 814, 1, 1. doi:10.1088/0004-637X/814/1/1
\bibitem[Lazzati et al.(2002)]{2002A&A...396L...5L} Lazzati, D., Rossi, E., Covino, S., et al.\ 2002, \aap, 396, L5. doi:10.1051/0004-6361:20021618
\bibitem[Lazzati \& Perna(2019)]{2019ApJ...881...89L} Lazzati, D. \& Perna, R.\ 2019, \apj, 881, 2, 89. doi:10.3847/1538-4357/ab2e06
\bibitem[Lei et al.(2013)]{2013ApJ...765..125L} Lei, W.-H., Zhang, B., \& Liang, E.-W.\ 2013, \apj, 765, 2, 125. doi:10.1088/0004-637X/765/2/125
\bibitem[Levan et al.(2024)]{2024Natur.626..737L} Levan, A.~J., Gompertz, B.~P., Salafia, O.~S., et al.\ 2024, \nat, 626, 8000, 737. doi:10.1038/s41586-023-06759-1
\bibitem[Li et al.(2016)]{2016ApJS..227....7L} Li, Y., Zhang, B., \& L{\"u}, H.-J.\ 2016, \apjs, 227, 1, 7. doi:10.3847/0067-0049/227/1/7
\bibitem[Li \& Paczy{\'n}ski(1998)]{1998ApJ...507L..59L} Li, L.-X. \& Paczy{\'n}ski, B.\ 1998, \apjl, 507, L59. doi:10.1086/311680
\bibitem[Liang et al.(2013)]{2013ApJ...774...13L} Liang, E.-W., Li, L., Gao, H., et al.\ 2013, \apj, 774, 1, 13. doi:10.1088/0004-637X/774/1/13
\bibitem[Liu et al.(2017)]{2017NewAR..79....1L} Liu, T., Gu, W.-M., \& Zhang, B.\ 2017, \nar, 79, 1. doi:10.1016/j.newar.2017.07.001
\bibitem[Liu et al.(2025)]{2025ApJ...988L..46L} Liu, X.-X., L{\"u}, H.-J., Chen, Q.-H., et al.\ 2025, \apjl, 988, 2, L46. doi:10.3847/2041-8213/adec83
\bibitem[L{\"u} et al.(2014)]{2014MNRAS.442.1922L} L{\"u}, H.-J., Zhang, B., Liang, E.-W., et al.\ 2014, \mnras, 442, 3, 1922. doi:10.1093/mnras/stu982
\bibitem[L{\"u} et al.(2010)]{2010ApJ...725.1965L} L{\"u}, H.-J., Liang, E.-W., Zhang, B.-B., et al.\ 2010, \apj, 725, 2, 1965. doi:10.1088/0004-637X/725/2/1965
\bibitem[L{\"u} \& Zhang(2014)]{2014ApJ...785...74L} L{\"u}, H.-J. \& Zhang, B.\ 2014, \apj, 785, 1, 74. doi:10.1088/0004-637X/785/1/74
\bibitem[L{\"u} et al.(2022)]{2022ApJ...931L..23L} L{\"u}, H.-J., Yuan, H.-Y., Yi, T.-F., et al.\ 2022, \apjl, 931, 2, L23. doi:10.3847/2041-8213/ac6e3a
\bibitem[L{\"u} et al.(2017)]{2017ApJ...835..181L} L{\"u}, H.-J., Zhang, H.-M., Zhong, S.-Q., et al.\ 2017, \apj, 835, 2, 181. doi:10.3847/1538-4357/835/2/181
\bibitem[L{\"u} et al.(2018)]{2018ApJ...862..130L} L{\"u}, H.-J., Lan, L., Zhang, B., et al.\ 2018, \apj, 862, 2, 130. doi:10.3847/1538-4357/aacd03
\bibitem[MacFadyen \& Woosley(1999)]{1999ApJ...524..262M} MacFadyen, A.~I. \& Woosley, S.~E.\ 1999, \apj, 524, 1, 262. doi:10.1086/307790
\bibitem[Malesani et al.(2004)]{2004ApJ...609L...5M} Malesani, D., Tagliaferri, G., Chincarini, G., et al.\ 2004, \apjl, 609, 1, L5. doi:10.1086/422684
\bibitem[Meszaros \& Rees(1993)]{1993ApJ...405..278M} Meszaros, P. \& Rees, M.~J.\ 1993, \apj, 405, 278. doi:10.1086/172360
\bibitem[Metzger et al.(2010)]{2010MNRAS.406.2650M} Metzger, B.~D., Mart{\'\i}nez-Pinedo, G., Darbha, S., et al.\ 2010, \mnras, 406, 2650. doi:10.1111/j.1365-2966.2010.16864.x
\bibitem[Metzger et al.(2011)]{2011MNRAS.413.2031M} Metzger, B.~D., Giannios, D., Thompson, T.~A., et al.\ 2011, \mnras, 413, 3, 2031. doi:10.1111/j.1365-2966.2011.18280.x
\bibitem[Metzger \& Piro(2014)]{2014MNRAS.439.3916M} Metzger, B.~D. \& Piro, A.~L.\ 2014, \mnras, 439, 4, 3916. doi:10.1093/mnras/stu247
\bibitem[Metzger(2019)]{2019LRR....23....1M} Metzger, B.~D.\ 2019, Living Reviews in Relativity, 23, 1, 1. doi:10.1007/s41114-019-0024-0
\bibitem[Modjaz et al.(2006)]{2006ApJ...645L..21M} Modjaz, M., Stanek, K.~Z., Garnavich, P.~M., et al.\ 2006, \apjl, 645, 1, L21. doi:10.1086/505906
\bibitem[M{\'e}sz{\'a}ros \& Rees(1997)]{1997ApJ...476..232M} M{\'e}sz{\'a}ros, P. \& Rees, M.~J.\ 1997, \apj, 476, 1, 232. doi:10.1086/303625
\bibitem[M{\'e}sz{\'a}ros et al.(1998)]{1998ApJ...499..301M} M{\'e}sz{\'a}ros, P., Rees, M.~J., \& Wijers, R.~A.~M.~J.\ 1998, \apj, 499, 1, 301. doi:10.1086/305635
\bibitem[Nakar \& Piran(2017)]{2017ApJ...834...28N} Nakar, E. \& Piran, T.\ 2017, \apj, 834, 1, 28. doi:10.3847/1538-4357/834/1/28
\bibitem[Nakar et al.(2003)]{2003JCAP...10..005N} Nakar, E., Piran, T., \& Waxman, E.\ 2003, \jcap, 2003, 10, 005. doi:10.1088/1475-7516/2003/10/005
\bibitem[Nathanail et al.(2021)]{2021MNRAS.502.1843N} Nathanail, A., Gill, R., Porth, O., et al.\ 2021, \mnras, 502, 2, 1843. doi:10.1093/mnras/stab115
\bibitem[Nugent et al.(2022)]{2022ApJ...940...57N} Nugent, A.~E., Fong, W.-F., Dong, Y., et al.\ 2022, \apj, 940, 1, 57. doi:10.3847/1538-4357/ac91d1
\bibitem[Palmer et al.(2025)]{2025GCN.39471....1P} Palmer, D.~M., Barthelmy, S.~D., Caputo, R., et al.\ 2025, GCN, 39471, 1. 
\bibitem[Palmerio et al.(2025)]{2025GCN.39418....1P} Palmerio, J.~T., Saccardi, A., Rayson, B., et al.\ 2025, GCN, 39418, 1.
\bibitem[Peng et al.(2005)]{2005ApJ...626..966P} Peng, F., K{\"o}nigl, A., \& Granot, J.\ 2005, \apj, 626, 2, 966. doi:10.1086/430045
\bibitem[Pian et al.(2006)]{2006Natur.442.1011P} Pian, E., Mazzali, P.~A., Masetti, N., et al.\ 2006, \nat, 442, 7106, 1011. doi:10.1038/nature05082
\bibitem[Popham et al.(1999)]{1999ApJ...518..356P} Popham, R., Woosley, S.~E., \& Fryer, C.\ 1999, \apj, 518, 1, 356. doi:10.1086/307259
\bibitem[Qin et al.(2013)]{2013ApJ...763...15Q} Qin, Y., Liang, E.-W., Liang, Y.-F., et al.\ 2013, \apj, 763, 1, 15. doi:10.1088/0004-637X/763/1/15
\bibitem[Racusin et al.(2008)]{2008Natur.455..183R} Racusin, J.~L., Karpov, S.~V., Sokolowski, M., et al.\ 2008, \nat, 455, 7210, 183. doi:10.1038/nature07270
\bibitem[Rastinejad et al.(2022)]{2022Natur.612..223R} Rastinejad, J.~C., Gompertz, B.~P., Levan, A.~J., et al.\ 2022, \nat, 612, 7939, 223. doi:10.1038/s41586-022-05390-w
\bibitem[Rees \& Meszaros(1994)]{1994ApJ...430L..93R} Rees, M.~J. \& Meszaros, P.\ 1994, \apjl, 430, L93. doi:10.1086/187446
\bibitem[Ren et al.(2024)]{2024ApJ...962..115R} Ren, J., Wang, Y., \& Dai, Z.-G.\ 2024, \apj, 962, 2, 115. doi:10.3847/1538-4357/ad1bcd
\bibitem[Rezzolla et al.(2011)]{2011ApJ...732L...6R} Rezzolla, L., Giacomazzo, B., Baiotti, L., et al.\ 2011, \apjl, 732, L6. doi:10.1088/2041-8205/732/1/L6
\bibitem[Rhoads(1997)]{1997ApJ...487L...1R} Rhoads, J.~E.\ 1997, \apjl, 487, 1, L1. doi:10.1086/310876
\bibitem[Ricci \& Troja(2025)]{2025GCN.39433....1R} Ricci, R. \& Troja, E.\ 2025, GCN, 39433, 1. 
\bibitem[Rossi et al.(2022)]{2022ApJ...932....1R} Rossi, A., Rothberg, B., Palazzi, E., et al.\ 2022, \apj, 932, 1, 1. doi:10.3847/1538-4357/ac60a2
\bibitem[Rossi et al.(2002)]{2002MNRAS.332..945R} Rossi, E., Lazzati, D., \& Rees, M.~J.\ 2002, \mnras, 332, 4, 945. doi:10.1046/j.1365-8711.2002.05363.x
\bibitem[Rowlinson et al.(2013)]{2013MNRAS.430.1061R} Rowlinson, A., O'Brien, P.~T., Metzger, B.~D., et al.\ 2013, \mnras, 430, 1061. doi:10.1093/mnras/sts683
\bibitem[Ryan et al.(2020)]{2020ApJ...896..166R} Ryan, G., van Eerten, H., Piro, L., et al.\ 2020, \apj, 896, 2, 166. doi:10.3847/1538-4357/ab93cf
\bibitem[Ryan et al.(2024)]{2024ApJ...975..131R} Ryan, G., van Eerten, H., Troja, E., et al.\ 2024, \apj, 975, 1, 131. doi:10.3847/1538-4357/ad6a14
\bibitem[Salvaggio et al.(2025)]{2025GCN.39414....1S} Salvaggio, C., Ferro, M., Williams, M.~A., et al.\ 2025, GCN, 39414, 1.
\bibitem[Sari et al.(1999)]{1999ApJ...519L..17S} Sari, R., Piran, T., \& Halpern, J.~P.\ 1999, \apjl, 519, 1, L17. doi:10.1086/312109
\bibitem[Sari et al.(1998)]{1998ApJ...497L..17S} Sari, R., Piran, T., \& Narayan, R.\ 1998, \apjl, 497, 1, L17. doi:10.1086/311269
\bibitem[Sarin et al.(2024)]{2024MNRAS.531.1203S} Sarin, N., H{\"u}bner, M., Omand, C.~M.~B., et al.\ 2024, \mnras, 531, 1, 1203. doi:10.1093/mnras/stae1238
\bibitem[Troja et al.(2022)]{2022Natur.612..228T} Troja, E., Fryer, C.~L., O'Connor, B., et al.\ 2022, \nat, 612, 7939, 228. doi:10.1038/s41586-022-05327-3
\bibitem[Troja et al.(2025)]{2025GCN.39491....1T} Troja, E., Becerra, R.~L., Zhang, W.~J., et al.\ 2025, GCN, 39491, 1. 
\bibitem[Usov(1992)]{1992Natur.357..472U} Usov, V.~V.\ 1992, \nat, 357, 6378, 472. doi:10.1038/357472a0
\bibitem[Wang et al.(2024a)]{2024ApJS..273...17W} Wang, H., Dastidar, R.~G., Giannios, D., et al.\ 2024a, \apjs, 273, 1, 17. doi:10.3847/1538-4365/ad4d9d
\bibitem[Wang et al.(2024b)]{2024MNRAS.527.5166W} Wang, H., Beniamini, P., \& Giannios, D.\ 2024b, \mnras, 527, 3, 5166. doi:10.1093/mnras/stad3560
\bibitem[Wang et al.(2025)]{2025ApJ...990..110W} Wang, H., Zhou, H., Fan, Y.-Z., et al.\ 2025, \apj, 990, 2, 110. doi:10.3847/1538-4357/adf1a2
\bibitem[Wang et al.(2026)]{2026JHEAp..5000490W} Wang, Y., Chen, C., \& Zhang, B.\ 2026, Journal of High Energy Astrophysics, 50, 100490. doi:10.1016/j.jheap.2025.100490
\bibitem[Watson et al.(2025)]{2025GCN.39397....1W} Watson, A.~M., Angulo, C., Lee, W.~H., et al.\ 2025, GCN, 39397, 1.
\bibitem[Wu et al.(2005)]{2005MNRAS.357.1197W} Wu, X.~F., Dai, Z.~G., Huang, Y.~F., et al.\ 2005, \mnras, 357, 4, 1197. doi:10.1111/j.1365-2966.2005.08685.x
\bibitem[Yang et al.(2015)]{2015NatCo...6.7323Y} Yang, B., Jin, Z.-P., Li, X., et al.\ 2015, Nature Communications, 6, 7323. doi:10.1038/ncomms8323
\bibitem[Yang et al.(2022)]{2022Natur.612..232Y} Yang, J., Ai, S., Zhang, B.-B., et al.\ 2022, \nat, 612, 7939, 232. doi:10.1038/s41586-022-05403-8
\bibitem[Yang et al.(2024)]{2024Natur.626..742Y} Yang, Y.-H., Troja, E., O'Connor, B., et al.\ 2024, \nat, 626, 8000, 742. doi:10.1038/s41586-023-06979-5
\bibitem[Yang et al.(2024)]{2024MNRAS.535.2482Y} Yang, Z., L{\"u}, H.-J., Yang, X., et al.\ 2024, \mnras, 535, 3, 2482. doi:10.1093/mnras/stae2496
\bibitem[Yu et al.(2013)]{2013ApJ...776L..40Y} Yu, Y.-W., Zhang, B., \& Gao, H.\ 2013, \apjl, 776, 2, L40. doi:10.1088/2041-8205/776/2/L40
\bibitem[Yuan et al.(2021)]{2021ApJ...912...14Y} Yuan, Y., L{\"u}, H.-J., Yuan, H.-Y., et al.\ 2021, \apj, 912, 1, 14. doi:10.3847/1538-4357/abedb1
\bibitem[Zhang et al.(2021)]{2021NatAs...5..911Z} Zhang, B.-B., Liu, Z.-K., Peng, Z.-K., et al.\ 2021, Nature Astronomy, 5, 911. doi:10.1038/s41550-021-01395-z
\bibitem[Zhang(2018)]{2018pgrb.book.....Z} Zhang, B.\ 2018, The Physics of Gamma-Ray Bursts by Bing Zhang. ISBN: 978-1-139-22653-0. Cambridge Univeristy Press, 2018. doi:10.1017/9781139226530
\bibitem[Zhang et al.(2009)]{2009ApJ...703.1696Z} Zhang, B., Zhang, B.-B., Virgili, F.~J., et al.\ 2009, \apj, 703, 2, 1696. doi:10.1088/0004-637X/703/2/1696
\bibitem[Zhang \& M{\'e}sz{\'a}ros(2002a)]{Zhang2002a} Zhang, B. \& M{\'e}sz{\'a}ros, P.\ 2002a, \apj, 571, 2, 876. doi:10.1086/339981
\bibitem[Zhang \& M{\'e}sz{\'a}ros(2002b)]{Zhang2002b} Zhang, B. \& M{\'e}sz{\'a}ros, P.\ 2002b, \apj, 566, 2, 712. doi:10.1086/338247
\bibitem[Zhang et al.(2004)]{2004ApJ...601L.119Z} Zhang, B., Dai, X., Lloyd-Ronning, N.~M., et al.\ 2004, \apjl, 601, 2, L119. doi:10.1086/382132
\bibitem[Zhang(2006)]{2006Natur.444.1010Z} Zhang, B.\ 2006, \nat, 444, 7122, 1010. doi:10.1038/4441010a
\bibitem[Zhang \& M{\'e}sz{\'a}ros(2001)]{2001ApJ...552L..35Z} Zhang, B. \& M{\'e}sz{\'a}ros, P.\ 2001, \apjl, 552, 1, L35. doi:10.1086/320255


\end{thebibliography}
\end{document}